


\input tcmjnl
\input eqnorder \referencefile

\title{ \bf
TOWARDS AN EXPERIMENTAL MEASURE OF THE NUMBER OF METASTABLE STATES
IN SPIN-GLASSES ?
 }

\author{J.P. Bouchaud}
\affil{TCM Group, Cavendish Laboratory,
Madingley Road, Cambridge CB3 0HE, UK and Service
de Physique de l'Etat Condens\'e, Orme des Merisiers,
CEA-Saclay, 91 191 Gif s/ Yvette CEDEX}
\author{E. Vincent, J. Hammann}
\affil{ Service
de Physique de l'Etat Condens\'e, Orme des Merisiers, CEA-Saclay, 91
191 Gif s/ Yvette CEDEX }
\abstract {
We show that an 'ergodic' time scale can be extracted from the
aging experiments on spin-glasses. This is the time for which a
significant fraction of the metastable states have been visited.
Experiments suggest that the number of metastable states per
independent subsystem is $\simeq 10^{12}-10^{14}$. For our
insulating spin-glass (CdCr$_{1.7}$In$_{0.3}$S$_4$) this number doubles when
the
temperature is lowered by $1 K$, but is roughly constant for the metallic
sample (Ag:Mn 2.6$\%$).

}


\endtopmatter
Aging experiments in spin-glasses
[\cite{exp1},\cite{exp2},\cite{exp3}] offer a unique way of
``scanning'' the phase space structure of complex systems, and a
detailed understanding of these experiments will probably offer a
key for this long debated problem
[\cite{BY},\cite{MP},\cite{FH},\cite{Russes}]. Very recently, one of
us proposed [\cite{JPB}] a phenomenological theory of aging based on
the assumption that energy barriers are distributed exponentially, as
suggested by mean field solutions of the spin-glass problem. The
probability to find the system in a given metastable state with
lifetime $\tau$ was shown to be `non-normalisable' in the spin-glass
phase - which means, more precisely, that the experimental time
itself is  needed to provide a long time cut-off and to give a
meaning to this probability distribution (see also
[\cite{Fei},\cite{Russes},\cite{Cohen}]).  The explicit
appearance of the `waiting-time' $t_w$ in the probability distribution
 implies that all
observables will depend on $t_w$ - this is the aging phenomenon.
Simple arguments, building upon these ideas, show that -say in
thermoremanent magnetisation relaxation (TRM) experiments - the
magnetisation decays as $f({t\over t_w})$, where $t$ is the time
after the switch off of the magnetic field, $t_w$ is the `waiting
time' during which the field is on, and $f(u)$ is a function which
behaves as a `stretched exponential' for small $u$, and as a power
law for large $u$ [\cite{JPB}]. This prediction is in very good
agreement with experiments over nearly five time decades for a {\it
given} $t_w$. However, curves for different $t_w$ fail to collapse to
the same curve when plotted versus the variable ${t\over t_w}$ (see
fig. 1-a): there is a systematic deviation, which suggests that the
relaxation of older systems is `faster' than it should be (although
of course, {\it when plotted versus} $t$, the older the system, the
slower the relaxation). In order to find  a
unique `Master' curve, the authors of [\cite{exp2}] proposed to take
into account this systematic shift by introducing a
relaxation time distribution of the form $\rho\left(\tau
\over (t+t_w)^\mu \right)$, where $\mu$ is around $0.9$.
This allows to rescale all data quite satisfactorily
as a
function of a reduced time variable which is equal to $t\over
t_w^\mu$ in the $t \ll t_w$ limit.  From a
fundamental point of view this procedure has an intriguing
consequence:  on dimensional grounds, the scaling variable should
really be written as ${t \over \tau^{*1-\mu} t_w^\mu}$. The
value of the time scale $\tau^*$ can be extracted from the
experiments and is extremely large [\cite{exp2}]: $\simeq 10^{20}$ seconds !
\par
The aim of this letter is to show that this systematic deviation
from a simple ${t \over t_w}$ scaling can be explained within the
framework of the theory proposed in [\cite{JPB}], {\it provided one
takes into account the finite number of available metastable
states}. A finite but quite long time scale indeed appears
naturally. \par In short, the idea is the following:  a finite size
system will eventually find the `deepest trap' in its phase space,
which corresponds to the equilibrium state [\cite{JPB}]. This will
take a long (but finite) time $t_{erg}$; when the waiting time $t_w$
exceeds this `ergodic' time $t_{erg}$, aging is `interrupted'
because the  phase space has been faithfully probed. Beyond this
time scale, conventional stationnary dynamics resume.
A comparison of this {\it equilibrium} dynamics with the
numerous available results [\cite{BY}], like
e.g. a.c. susceptibility experiments at not too low frequency, is
beyond the scope of the present paper, which mainly addresses the
question of aging in TRM experiments. \par
 Experimentally, the samples investigated can be
thought of as collections of independent subsystems, each of which
evolving within its own complicated phase space. There will thus be
a {\it distribution} of ergodic times - some shorter and some
longer. The systematic deviation from a perfect ${t\over t_w}$ scaling is the
consequence of the partial `death' of the system, i.e. that aging is
interrupted in a certain fraction of subsystems. The model proposed
in [\cite{JPB}] allows to quantify this effect, and in principle to
extract from the spin-glass data a typical ergodic time of the
system - which is in turn related to the number of available
metastable states. In fact, this 'interrupted aging' phenomenon has
probably been observed on glassy Charge Density Wave systems
[\cite{Bijla}], where the ergodic time scale appears to be much
shorter than in spin-glasses.\par Let us briefly recall the results
of ref. [\cite{JPB}]. An exponential distribution of barrier heights
(suggested by the mean-field models of spin-glasses
[\cite{REM},\cite{MP}]) induces a power-law distribution for the
lifetime of metastable states: $$\psi (\tau) = {\tau_0^x \over
\tau^{1+x}}\ \ \ (\tau \gg \tau_0)\eqno(1)$$ where $\tau_0$ is a
microscopic time scale ($\sim 10^{-12} $ sec.) and  $x$ is a
parameter  describing the structure of phase space. When $x$ is
between $0$ and $1$, Eq. (1) indicates that the average exploration
time is infinite: this leads to aging. For $x>1$, the average
exploration time is finite and conventional stationnary relaxation
occurs.\par If the total number of metastable states of a given
subsystem is $\cal S$, the ergodic time is of the order of the
longest $\tau$ drawn from distribution (1):   $t_{erg}({\cal S})
\simeq \tau_0 {\cal S}^{1 \over x}$. The relaxation of the
magnetisation $m(t,t_w,t_{erg}({\cal S}))$ after a given waiting
time $t_w$ is found to be,  for $t_w \ll t_{erg}({\cal S})$,
$f({t\over t_w})$ with $f(u)= \exp -\left[{\gamma_I \over 1-x}
u^{1-x}\right]$ for $u \ll 1$, and $f(u) \propto u^{-\gamma_{II}}$
for $u \gg 1$ [\cite{JPB}]. $\gamma_I$ is a number quantifying the
tendency of the system to loose its magnetisation rather than
continue to evolve within magnetized states. $\gamma_{II}$ is
proportional to $\gamma_{I}$, with an $x$ dependent coefficient
(which was, for simplicity, taken to be 1 in [\cite{JPB}] - see
below).  For longer times $t_w \gg t_{erg}({\cal S})$,
$m(t,t_w,t_{erg}({\cal S}))$ becomes a function of ${t \over
t_{erg}({\cal S})}$. This function is in fact very nearly the  same
$f$ [\cite{JPB}] (and the small difference does not affect the
following conclusions: only its {\it monotonic} character is of
importance).
 Hence, the
magnetisation of a given subsystem can be written in a compact form as: $$
m(t,t_w,t_{erg}({\cal S})) =  m_0({\cal S}) f\left({t \over
t^*[{t+t_w\over t_{erg}({\cal S})}]} \right) \eqno(2) $$ with
$t^*[.]=t_w$ for $t+t_w \ll t_{erg}({\cal S})$ and $t^*[.]=t_{erg}({\cal
S})$ for $t+t_w \gg t_{erg}({\cal S})$. In words, Eq. (2) means that once
equilibrium is achieved ($t+t_w \gg t_{erg}({\cal S})$),
the magnetisation is much
smaller than its `short' time extrapolation, $f({t\over t_w})$.\par
 Now, assuming that the
number of metastable states varies from one subsystem to another
according to a distribution ${\cal P}({{\cal S}\over {\cal S}_0})$
(${\cal S}_0$ is thus a 'typical' number of metastable states per
subsystem), the total (observed) magnetisation $M(t,t_w)$ is
obtained as the average over $\cal P$ of eq. (2). With these only
assumptions, one finally finds:
$${M(t,t_w) \over M_0} = f({t \over t_w}) \left(1 - {\cal F}
\left[({t+t_w \over  t_{erg}^0})^x\right]\right) \eqno(3) $$
where $ t_{erg}^0
\equiv \tau_0 S_0^{1 \over x}$ is the `ergodic time' of the
system, and ${\cal F}[.]$ is an increasing function which depends on the
choice of $m_0({\cal S})$ and ${\cal P}$. Before taking a specific
example, it should be noted that eq. (3) qualitatively agrees with
the experimental trend: older systems indeed relax `faster' {\it
when plotted versus } ${t \over t_w}$.\par  The simplest assumption
is that $m_0$ and $\cal P$ behave regularly when their arguments are
small. In this case, one finds, for $t+t_w \ll t_{erg}^0$
 ${\cal F} (u) = A u + Bu^2 +...$, where $A$ is a
constant which can be reabsorbed in the definition of $t_{erg}^0$ [\cite{Rq1}].
We show in fig. 1-b the same data as in fig. 1-a, but with the
`interrupted aging' correction, eq. (3). The only free parameters
are the ergodic time of the system $t_{erg}^0({\cal S})$ and the
value of $B$ (the value of $x$ is fixed by fitting the `youngest' curve -
see [\cite{Rq1}]). We have
however imposed the same value of $B=-3$ (i.e. the same functional form for
${\cal
P}$) for all temperatures and samples. The collapse of the curves is quite
good in all cases (see fig 1-b). The corresponding values of the parameters
$x,\gamma_{I},\gamma_{II}$, and of the `complexity' $\Omega \equiv x
\log_{10} ({t_{erg}^0 \over \tau_0})$ are given in Table I and II,
for different temperatures and two different samples: Ag:Mn 2.6$\%$
(AgMn), which is metallic and CdCr$_{1.7}$In$_{0.3}$S$_4$ (CrIn)
which is an insulating  spin glass. It is interesting to see that
both samples   lead to similar conclusions, although noticable
differences appear: \par a) The ergodic time $t_{erg}^0$ is of the
order of $10^6$ seconds, and thus, taking $\tau_0 \simeq 10^{-12}$
sec., the number of metastable states per 'subsystem' is found to be
$\simeq 10^{12}-10^{14}$ (see Table I,II).  This number is however
not easy to discuss since we do not know precisely what these
'subsystems' are. We believe that they are magnetically disconnected
regions (like grains) the size of which being determined by the
sample preparation, and thus is temperature independent. We however
think that numerical simulations [\cite{Rieger}] and/or experiments
on mesoscopic spin-glasses [\cite{Weiss}] are really needed to
understand how the ergodic time $t_{erg}^0$ is related to the number
of spins contained in a subsystem. \par b) As shown in figure 2,
the number of metastable states of the insulating sample
(CrIn) tends to increase when the temperature is lowered
(it roughly doubles when
the temperature is lowered by 1 Kelvin). This is expected
from the work of Bray and Moore on the SK model [\cite{Bray}].
However, in contrast to the SK model, {\it this number does not
vanish continuously at the spin-glass transition} $T_g$. Such a
behaviour was found in `Potts glass' models [\cite{TK}] and in
models with p$(>2)$- spin interactions [\cite{REM},\cite{Rieger2}].
The number of metastable states of the metallic sample (AgMn), on
the other hand, seems to be roughly temperature independent. This
difference in behaviour between the two kind of spin-glasses may
reflect the different types of microscopic interactions (exchange
versus RKKY).
 \par  c) The parameter $x$
grows towards $1$ when $T \longrightarrow T_g$ (at least for AgMn,
but other experiments on CrIn showed the same tendency
[\cite{exp2}]). This behaviour was also found in a very clear way in
ref.[\cite{Hoo}], where the initial decay of the magnetisation was
also fitted by a stretched exponential. Our
results ($\Omega(T_g)$ finite and $x(T_g)\simeq 1$) suggest that the
spin-glass transition is, in these samples, much closer to the
`Random Energy Model' scenario [\cite{REM}], than to the `SK'
scenario where $\Omega(T_g)=0$ and $x(T_g)=0$ [\cite{MP}].\par d)
Finally, one can see that the simplifying hypothesis $\gamma_I =
\gamma_{II}$ is quite accurate. In fact, a closer look at equations
(8-10) of ref. [\cite{JPB}] reveals that the ratio ${\gamma_{II}
\over \gamma_I}$ increases with $x$ (and diverges for $x
\longrightarrow 1$), which is  consistent with the experimental
trend (see Table I). \par
 Let
us summarize our findings: quantifying the idea that aging stops when
the exploration of phase space is completed, we have proposed a way
to rescale relaxations obtained for different waiting times on a
unique 'Master' curve. This allows to extract from the experiments a
'very long', temperature dependent, time scale beyond which aging
ceases and the nature of relaxation changes. This is interesting
from two standpoints: from a fundamental point of view, it allows -
to our knowledge, for the first time -  to estimate the number of
metastable states in a complex system, and its temperature
dependence. Perhaps more importantly, it allows, from a practical
point of view, to extract from relatively short time measurements
an ergodic time beyond which the relaxation changes qualitatively: we
have in mind here the aging experiments on polymer glasses, where the
mechanical properties are very similar to those discussed here
[\cite{Struik},\cite{Perez}]. It can be of crucial importance to
know when these mechanical properties will appreciably change. It
would thus be very interesting to apply the ideas developed here to
other (molecular or metallic) glasses. \vskip 0.5cm
{\it Acknowledgments} We wish to express warm gratitude towards
M. Ocio, who gave us access to his data on AgMn, and for enlightening
critical discussions.
\vskip 0.5cm
{\it Figure Captions}
\par Fig 1: 1-a. TRM relaxation in Ag:Mn 2.6$\%$ at 9K, as a function
of $\log({t\over t_w})$, for $t_w$= 300, 1000, 3000, 10000 and 30000
seconds. The systematic shift towards faster relaxation as $t_w$
increases is clearly seen. Insert: Sketch of the experimental
TRM procedure. \par 1-b: Same data as in Fig. 1-a, but with the
'interrupted aging' correction, eq. (3), expanded to second order in
$t+t_w \over t_{erg}^0$. The error bar is an estimate of the
systematic uncertainty in the vertical axis. \par Fig 2:
Evolution of the 'complexity' $\Omega \equiv x \log_{10} {t_{erg}^0
\over \tau_0}$ with reduced temperature $T \over T_g$ for both AgMn
and CrIn. Note that the two systems evolve differently with
temperature.  Insert: Evolution of the complexity per spin versus
reduced temperature for the p-spin model, with, from bottom to top:
p=2 (SK model), 3,4,5 and 6 (p=$\infty$ is the Random Energy Model).
Note that $\Omega$ is discontinuous at $T_g$ as soon as $p \geq 3$.
 From ref. [\cite{Rieger2}].  \vskip 1cm {\it Table Captions}\par
Table 1: Parameters $x,\gamma_I$ and $\gamma_{II}$ and `complexity'
$\Omega$ for the Ag:Mn 2.6$\%$ sample [\cite{exp2}]. $x,\gamma_I$
were extracted by fitting the `initial' part of the TRM relaxation
($t \leq 0.3 t_w$), while $\gamma_{II}$ is obtained from the long
time part ($t \geq 2 t_w$) of the  relaxations.  Note that
$\Omega(T=10K)$ could not be estimated because of our lack of
precision in the vertical positionning of the relaxation curves.
This is a problem close to $T_g$ since the magnetisation falls
rapidly to small values.\par Table II: Same as in Table I, but for
the CdCr$_{1.7}$In$_{0.3}$S$_4$
 sample [\cite{exp2}]. Here, the initial part of the relaxation
was extended to $t \leq 0.5 t_w$. The
value of the parameters given here are compatible with those
obtained in ref. [\cite{JPB}], where the additionnal assumption
$\gamma_I \equiv \gamma_{II}$ was made.
 \par

\def\strut{\vrule height10pt width 0pt depth3.5pt}

\vskip 1cm

\vbox{\tabskip=0pt \offinterlineskip
\def\tablerule{\noalign{\hrule}}
\def\strut{\vrule height10pt width 0pt depth3.5pt}
\halign to 400pt {\strut#& \vrule#\tabskip=1em plus2em&
  \hfil#\hfil& \vrule#& \hfil#\hfil& \vrule#& \hfil#\hfil& \vrule#&
\hfil#\hfil& \vrule#&
  \hfil#\hfil& \vrule#\tabskip=0pt\cr\tablerule
&&\multispan9\hfil Table I: AgMn: $2.6 \%$ ($T_g$=10.4
K)\hfil&\cr\tablerule\tablerule
&&$T({\rm
K})$&&$x$&&$\gamma_I$&&$\gamma_{II}$&&$\Omega$&\cr\tablerule\tablerule
&&8&&0.65$\pm$0.05&&0.083$\pm$0.010&&0.090$\pm$0.003&&12.1$\pm$0.9
&\cr\tablerule
&&9&&0.65$\pm$0.05&&0.0124$\pm$0.015&&0.127$\pm$0.003&&11.9$\pm$0.9&
\cr\tablerule
&&9.5&&0.68$\pm$0.05&&0.153$\pm$0.020&&0.155$\pm$0.005&&12.2$\pm$0.9&
\cr\tablerule
&&10&&0.75$\pm$0.05&&0.221$\pm$0.030&&0.265$\pm$0.005&&-&\cr\tablerule}}
\vskip 1cm

\vbox{\tabskip=0pt \offinterlineskip
\def\tablerule{\noalign{\hrule}}
\def\strut{\vrule height10pt width 0pt depth3.5pt}
\halign to 400pt {\strut#& \vrule#\tabskip=1em plus2em&
  \hfil#\hfil& \vrule#& \hfil#\hfil& \vrule#& \hfil#\hfil& \vrule#&
\hfil#\hfil& \vrule#&
  \hfil#\hfil& \vrule#\tabskip=0pt\cr\tablerule
&&\multispan9\hfil Table II: CdCr$_{1.7}$In$_{0.3}$S$_4$ ($T_g$=16.7
K)\hfil&\cr\tablerule
&&$T({\rm
K})$&&$x$&&$\gamma_I$&&$\gamma_{II}$&&$\Omega$&\cr\tablerule\tablerule
&&10&&0.76$\pm$0.05&&0.059$\pm$0.008&&0.058$\pm$0.003&&14.2$\pm
$0.9&\cr\tablerule
&&12&&0.76$\pm$0.05&&0.100$\pm$0.01&&0.100$\pm$0.003&&14.0$\pm
$0.9&\cr\tablerule
&&14&&0.73$\pm$0.05&&0.200$\pm$0.02&&0.182$\pm$0.010&&13.0$\pm
$0.9&\cr\tablerule}
}

\vskip 1cm
\references {
\refis{exp1} L. Lundgren, P. Svedlindh, P. Nordblad, O. Beckmann, Phys. Rev.
Lett 51,
911 (1983),
P. Nordblad, L. Lundgren, P. Svedlindh, L. Sandlund, Phys. Rev. B 33, 645
(1988)
\refis{exp2} a) M. Alba, M. Ocio, J. Hammann, Europhys. Lett. 2, 45 (1986), J.
Phys.
Lett. 46 L-1101 (1985), M. Alba, J. Hammann, M. Ocio, Ph. Refregier, J. Appl.
Phys.
61, 3683 (1987). b) E. Vincent, J. Hammann, M. Ocio, p. 207 in
"Recent Progress in Random Magnets", D.H. Ryan Editor, (World
Scientific Pub. Co. Pte. Ltd, Singapore 1992) \refis{exp3} J. Hammann,
M. Lederman, M. Ocio, R. Orbach, E. Vincent, Physica A 185, 278
(1992), F. Lefloch, J. Hammann, M. Ocio, E. Vincent, Europhys.
Lett 18, 647 (1992)
 \refis{BY} K. Binder, A.P. Young, Rev. Mod. Phys. 58, 801 (1986)
\refis{MP} M. M\'ezard, G. Parisi, M.A. Virasoro, ``Spin Glass Theory and
Beyond'',
(World Scientific, Singapore 1987)
 \refis{FH} D.S  Fisher, D.A. Huse , Phys. Rev. Lett 56, 1601 (1986), Phys.
Rev. B 38,
373 (1988)
\refis{Russes}
V.S. Dotsenko, M. V. Feigelmann, L.B. Ioffe, Spin-Glasses and related problems,
Soviet
Scientific Reviews, vol. 15 (Harwood, 1990)
\refis{JPB} J.P. Bouchaud, J. Physique I France 2, 1705 (1992)
\refis{Fei} M. V. Feigelmann, V. Vinokur, J. Phys. France 49, 1731 (1988)
\refis{Cohen} F. Bardou, J.P. Bouchaud, C. Cohen-Tannouji, O. Emile: ``A
Statistical
approach to dark resonances'', unpublished.
\refis{Bijla} K. Biljakovic, J.C. Lasjaunias, P. Monceau, F. L\'evy, Phys. Rev.
Lett.
67, 1902 (1991), K. Biljakovic, J.C. Lasjaunias, P. Monceau, Phys. Rev. Lett
62, 1512
(1989)
\refis{REM} B. Derrida, Phys. Rev. B 24,
2613 (1981)
\refis{Rq1} Since $t_{erg}^0$ turns out to be $\simeq 10^6$ seconds, one may
proceed `backwards' as follows. The `youngest' relaxation curve is such that
the
correction term in eq. (3) is negligible: this allows to identify the Master
curve
$f({t \over t_w})$. Now, plotting the {\it ratio} of -say- the oldest
relaxation curve
to the youngest versus the variable $(t+t_w)^x$ determines the function ${\cal
F}(u)$.
In the interval of $u$ which is probed, a fit of the form ${\cal F}(u) = u +
Bu^2$ is
very satisfactory.
 \refis{Rieger} H. Rieger, preprint, submitted to Phys.
Rev. Lett.
\refis{Weiss} G. B. Alers, M. B. Weissmann, N.E. Isrealoff, Phys.
Rev. B 46, 507 (1992),  M. B. Weissmann, N.E. Isrealoff,  G. B.
Alers,
Journal of Magn. Magn. Mat. 114, 87 (1992), and M. B. Weissmann,
Rev. Mod. Phys., July 1993 (to appear).
 \refis{Bray} A. J. Bray, M.
A. Moore, J. Phys. C 13, L469 (1980)
\refis{TK} D. Thirumalai, T. R. Kirkpatrick, Phys. Rev. B 36,
5388 (1987), T. R. Kirkpatrick, P. G. Wolynes, Phys. Rev. B 36, 8552
(1987)
\refis{Rieger2} H. Rieger, Phys. Rev. B 46, 14655 (1992)
\refis{Hoo} R. Hoogerbeets, W. L. Luo, R. Orbach, Phys. Rev. B 34, 1719 (1986)
. Note however that in this paper, $x$ was interpreted
as $1-x$. See [\cite{JPB}] for a discussion of this
point.
\refis{Struik} L.C.E. Struick, ``Physical Aging in Amorphous
Polymers and Other Materials'' (Elsevier, Houston, 1978)
\refis{Perez} J. Perez, `Physique et Mecanique des Polym\`eres
Amorphes', Technique et Documentation Lavoisier, Paris, 1992.

}

\endreferences

 \end